\documentclass[times,twocolumn,final]{elsarticle}

\usepackage{medima}
\usepackage{framed,multirow}

\usepackage{amssymb}
\usepackage{latexsym}

\usepackage{url}
\usepackage{xcolor}
\usepackage{hyperref}
\definecolor{newcolor}{rgb}{.8,.349,.1}

\usepackage{mathtools}
\usepackage{amsmath}
\DeclareMathOperator{\E}{\mathbb{E}}

\journal{Medical Image Analysis}

\begin{document}

\verso{B. Zhou \textit{et~al.}}

\begin{frontmatter}

\title{Anatomy-guided Multimodal Registration by Learning Segmentation without Ground Truth: Application to Intraprocedural CBCT/MR Liver Segmentation and Registration}

\author[1]{Bo \snm{Zhou}}
\ead{bo.zhou@yale.edu}
\author[1]{Zachary \snm{Augenfeld}}
\author[2]{Julius \snm{Chapiro}}
\author[4,5]{S. Kevin \snm{Zhou}}
\author[1,2]{Chi \snm{Liu}}
\author[1,2,3]{James S. \snm{Duncan}\corref{cor1}}
\cortext[cor1]{Corresponding author.}
\ead{james.duncan@yale.edu}

\address[1]{Department of Biomedical Engineering, Yale University, New Haven, CT, USA}
\address[2]{Department of Radiology and Biomedical Imaging, Yale School of Medicine, New
Haven, CT, USA}
\address[3]{Department of Electrical Engineering, Yale University, New Haven, CT, USA.}
\address[4]{School of Biomedical Engineering \& Suzhou Institute for Advanced Research, University of Science and Technology of China, Suzhou, China.}
\address[5]{Institute of Computing Technology, Chinese Academy of Sciences, Beijing, China.}







\begin{abstract}
Multimodal image registration has many applications in diagnostic medical imaging and image-guided interventions, such as Transcatheter Arterial Chemoembolization (TACE) of liver cancer guided by intraprocedural CBCT and pre-operative MR. The ability to register peri-procedurally acquired diagnostic images into the intraprocedural environment can potentially improve the intra-procedural tumor targeting, which will significantly improve therapeutic outcomes. However, the intra-procedural CBCT often suffers from suboptimal image quality due to lack of signal calibration for Hounsfield unit, limited FOV, and motion/metal artifacts. These non-ideal conditions make standard intensity-based multimodal registration methods infeasible to generate correct transformation across modalities. While registration based on anatomic structures, such as segmentation or landmarks, provides an efficient alternative, such anatomic structure information is not always available. One can train a deep learning-based anatomy extractor, but it requires large-scale manual annotations on specific modalities, which are often extremely time-consuming to obtain and require expert radiological readers. To tackle these issues, we leverage annotated datasets already existing in a source modality and propose an anatomy-preserving domain adaptation to segmentation network (APA2Seg-Net) for learning segmentation without target modality ground truth. The segmenters are then integrated into our anatomy-guided multimodal registration based on the robust point matching machine. Our experimental results on in-house TACE patient data demonstrated that our APA2Seg-Net can generate robust CBCT and MR liver segmentation, and the anatomy-guided registration framework with these segmenters can provide high-quality multimodal registrations.
\end{abstract}


\end{frontmatter}


\section{Introduction}
Primary liver cancer is the fourth most common cancer and the third most common cause of cancer-related mortality worldwide with incidence rates rising across the globe and especially in the United States and
Europe \citep{bray2018global}. Local image-guided therapies, such as Transcatheter Arterial Chemoembolization (TACE), are commonly used procedures that are performed in patients with intermediate to advanced stages as a palliative therapy option, capable of significantly prolonging patient survival \citep{pung2017role}. Most patients undergo multi-parametric multi-phasic contrast-enhanced MRI using gadolinium-enhanced T1 sequences both for diagnostic purposes as well as for the sake of therapy planning. This readily available multi-parametric information on tumor vascularity, size, location and even tissue properties is clinically underutilized for intra-procedural navigation primarily for technical reasons such as lack of practical image registration solutions. Intraprocedural navigation and targeting is instead achieved with serial, planar angiographic imaging as well as intra-procedural Cone-beam Computed Tomography (CBCT) imaging that provides a coarse cross-sectional dataset, which can then be used to map arterial supply of the tumor and allow for accurate catheter guidance and intra-procedural feedback. While CBCT utilizes an x-ray source and the high-resolution 2D flat panel detector enables fast 3D organ visualization during procedures, CBCT suffers from low contrast-to-noise ratio (CNR), narrow abdominal tissue dynamic range, limited field-of-view (FOV), and motion/metal-induced artifacts, making it challenging to directly visualize and localize targeted tumors \citep{tacher2015cone,pung2017role}. Therefore, multimodal image registration, i.e. mapping preoperative MR imaging and associated liver segmentations to intraprocedural CBCT is essential for accurate liver/tumor localization, targeting and subsequent drug delivery. Current workflows do not apply any quantitative measurements on the acquired CBCT images and the predominant technique is mere “gestalt” assessment of the images. Automatic multimodal registration is therefore highly desirable in image-guided interventional procedures.

\begin{figure*}[htb!]
\centering
\includegraphics[width=0.98\textwidth]{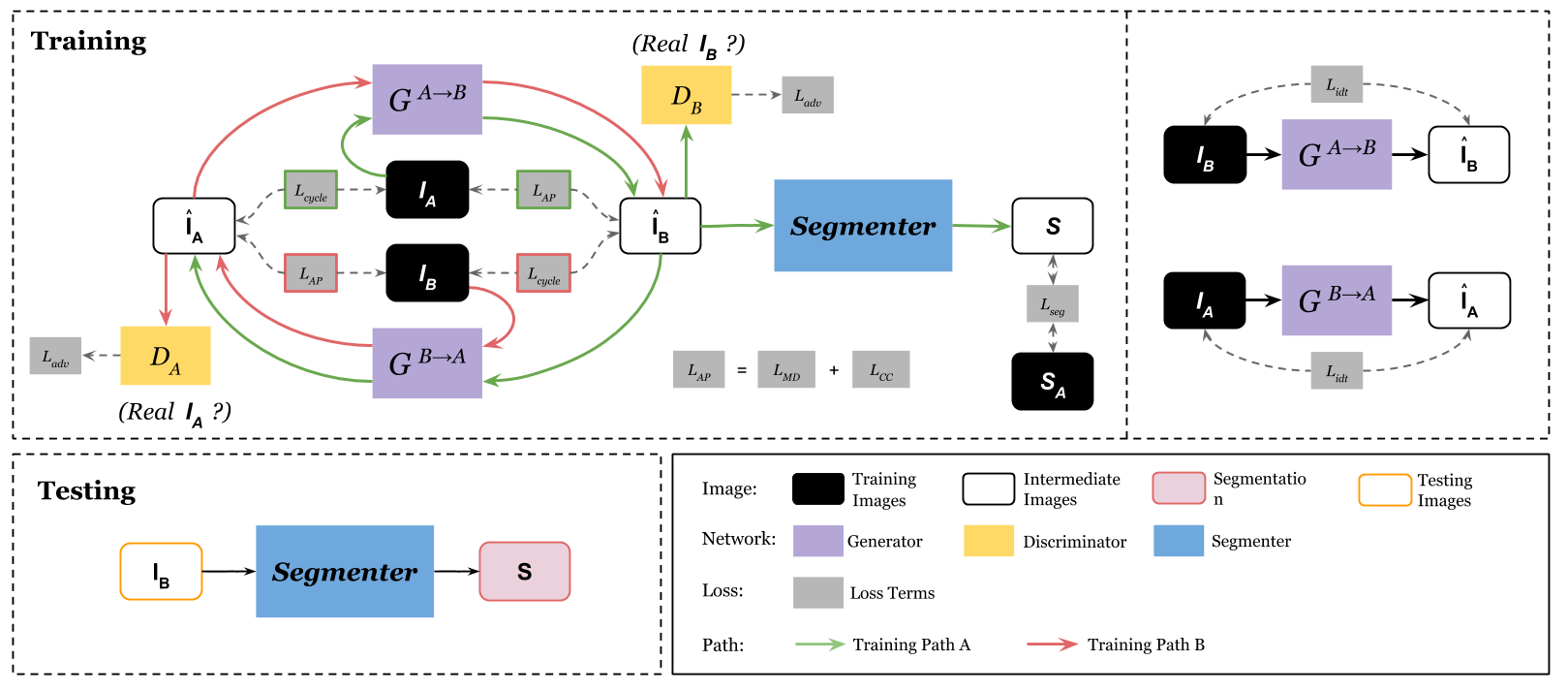}
\caption{Illustration of our \textbf{A}natomy-\textbf{P}reserving domain \textbf{A}daptation to \textbf{Seg}mentation Network (\textbf{APA2Seg-Net}). It consists of an anatomy-preserving domain adaptation network (left portion), and a segmenter for target domain. During test phase, the segmenter is extracted from APA2Seg-Net for predicting structural information, i.e., segmentation. }
\label{fig:network}
\end{figure*}

Previous multimodal/monomodal image registration algorithms can be categorized into two classes: conventional iterative based approaches \citep{wyawahare2009image,maes1997multimodality,avants2008symmetric,rohr2001landmark,heinrich2012mind} and deep learning based approaches \citep{hu2018weakly,qin2019unsupervised,lee2019image,zhou2020simultaneous,wang2020deepflash,arar2020unsupervised,mok2020fast}. Conventional approaches utilize iterative maximization of intensity similarity metrics, such as mutual information \citep{maes1997multimodality}, cross correlation \citep{avants2008symmetric} and difference in MIND \citep{heinrich2012mind}, to find the optimal registration transformation between images. If paired key points between the images are available, landmark-based thin-plate splines registration can be applied to estimate the transformation between images. Previously, \citet{al2015assessing,al2017usefulness} demonstrated the feasibility of temporomandibular joints MRI-CBCT registration via the above mentioned intensity-based and landmark-based registration methods. In the application of head and neck CBCT-CT registration, \citet{zhen2012ct,park2017deformable} proposed to integrate the intensity matching between CT and CBCT into conventional iterative based approaches for more accurate CT-CBCT registrations. However, the intensity matching approaches are suitable for either monomodal or multimodal with similar imaging physics and cannot be adapted to CBCT-MR registration. More recently, \citet{solbiati2018novel} proposed a two-stage registration for CBCT-CT liver registration, where manually annotated key points in the first stage are used for coarse alignment and conventional iterative registration based on mutual information is subsequently performed to refine the alignment.

With the recent advances in data driven learning~\citep{zhou2020review}, deep learning based methods have achieved comparable registration performance with a significantly higher inference speed. For monomodal registration, \citet{balakrishnan2019voxelmorph} proposed the first deep learning based registration method using a deep convolutional network to predict the registration transformation between monomodality images, called VoxelMorph. \citet{mok2020fast} further improved its registration performance by adding the symmetric diffeomorphic properties into the network design. Moreover, \citet{wang2020deepflash} developed a learning-based registration framework, called DeepFLASH, that utilizes low dimensional band-limited space for efficient transformation field computing. For multimodal registration, \citet{hu2018weakly} proposed to use the organ segmentations for weakly supervised training the transformation estimation network, where intensity-neutral supervision makes the multimodal registration feasible. However, accurate manual organ segmentation is required for their approach and thus limits its applications. As an alternative to this approach, \citet{qin2019unsupervised} proposed to estimate the non-rigid transformation from disentangled representation of multimodal image contents. There are also recent studies of multimodal image registration for natural images \citep{arar2020unsupervised}, where source image appearance is first translated to fix image appearance, and then previously established monomodal registration methods \citep{balakrishnan2019voxelmorph} are applied. Although all the above methods achieve impressive results, they are limited to multimodal registration with no occluded FOV, sufficiently wide intensity range, or organ segmentations. Those conditions are hardly satisfied in many image-guided intervention procedures, such as TACE. More recently, \citet{augenfelf2020automatic} proposed to use manual CBCT liver annotation to train a CBCT segmenter and register based on the predicted CBCT segmentation and manually annotated MR. While demonstrating the feasibility of registration in TACE, segmenting liver on intraprocedural image is not clinical routine and training such segmenter from limited annotation data impede the segmentation and registration performance. 

To tackle these issues, we present an anatomy-guided registration framework by learning segmentation without target modality ground truth. In previous works of learning segmentation without target modality, \citet{zhang2018task} proposed a two-step strategy, where they first use CycleGAN \citep{zhu2017unpaired} to adapt the target domain image to the domain with a well-trained segmenter, and then predict the segmentation on the adapted image. However, the segmentation performance relies on the image adaptation performance, thus the two-step process may prone to error aggregation. To improve the CycleGAN performance in medical imaging, \citet{zhang2018translating} suggested adding two segmenters as additional discriminators for generating shape-consistent image adaption results. However, ground-truth segmentation is required for both source and target domains. Recently, \citet{yang2020unsupervised} proposed to add the Modality Independent Neighborhood Descriptor (MIND) loss \citep{heinrich2012mind} in the CycleGAN to constrain image structure during the adaptation. Similarly, \citet{ge2019unpaired} proposed to incorporate correlation coefficient loss in the CyleGAN to constrain image structure. Both strategies demonstrated improvements in MR-to-CT translation. On the other hand, \citet{huo2018adversarial} proposed Syn2Seg-Net that merges CycleGAN with a segmentation network on the target domain output, such that the segmentation network is trained on the target domain without target domain ground truth. However, the training image of the segmentation network relies on high-quality adapted images from the CycleGAN part of SynSeg-Net. Without anatomy-preserving constraint during the adaptation, the image could be adapted to a target domain image with incorrect anatomical contents, and negatively impact the subsequent segmentation network's training. Inspired by \citet{huo2018adversarial} and with large-scale manual liver segmentation on conventional CT available from public dataset, such as LiTS \citep{bilic2019liver} and CHAOS \citep{kavur2020chaos}, we propose an anatomy-preserving domain adaptation to segmentation network (APA2Seg-Net) for learning segmentation without CBCT/MR ground truth. Specifically, we aim to use only conventional CT segmentation to train robust CBCT/MR segmenters in an anatomy-preserving unpaired fashion. The extracted anatomic information of CBCT/MR, i.e liver segmentations, guides our Robust Point Matching (RPM) to estimate the multimodal registration transformation. Our experimental results on TACE patients demonstrate that our APA2Seg-Net based registration framework allows us to get robust target modality segmenters without ground truth, and enables accurate multimodal registration. Our code is available at \href{https://github.com/bbbbbbzhou/APA2Seg-Net}{https://github.com/bbbbbbzhou/APA2Seg-Net}.

\section{Methods}
We propose a novel two-stage multimodal registration framework for mapping pre-operative MR to intraprocedural CBCT for liver image-guided interventions. In the first stage, our APA2Seg-Net is trained with 3 sources of images: paired conventional CT with liver segmentation and unpaired CBCT/MR, such that CBCT/MR segmenters can be extracted from our APA2Seg-Net for outputting the anatomic information. In the second stage, we extract the surface points of the outputted CBCT and MR segmentations, and input them into our RPM machine to predict MR to CBCT transformation. Finally, the transformation is applied to the pre-operative MR and the associated labels to register to the intraprocedural CBCT. The details are discussed in following sections.

\begin{figure*}[htb!]
\centering
\includegraphics[width=0.94\textwidth]{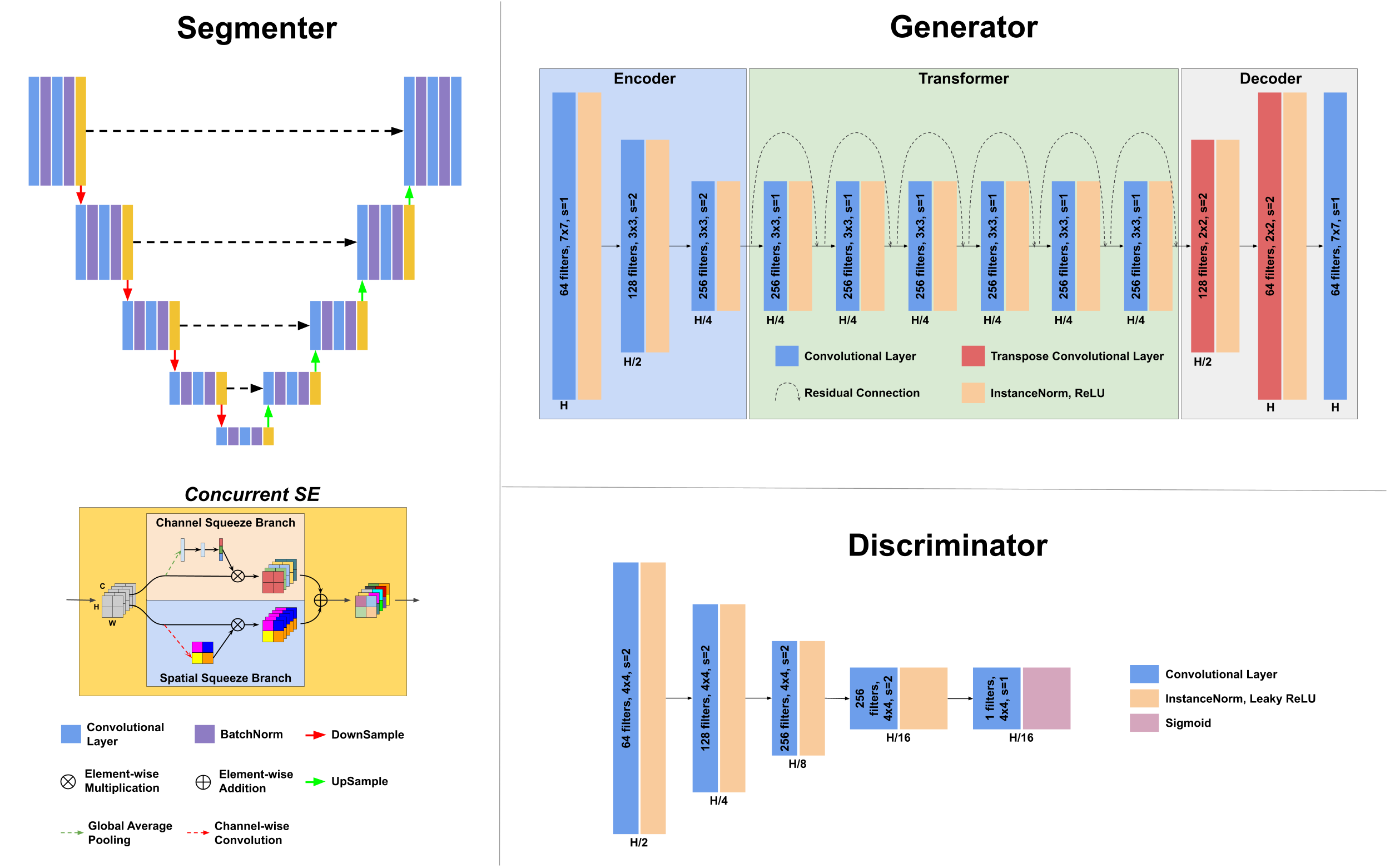}
\caption{Illustration of our segmenter, generator, and discriminator network structures in the APA2Seg-Net. 5-level U-Net with the concurrent squeeze and excitation module is used for our segmenter. An autoencoder with multiple residual bottleneck is used for our generator. The feature size shrinks in the encoder phase, stays constant in the transformer phase, and expands again in the decoder phase. The feature size of the layer outputs is listed below it, in terms of the input image size, H. On each layer is listed the number of filters, the size of those filters, and the stride.}
\label{fig:network_details}
\end{figure*}

\subsection{Anatomy-preserving Adaptation to Segmentation Network}
Our APA2Seg-Net consists of two parts - an anatomy-preserving domain adaptation network (APA-Net) and a segmentation network. The architecture and training/test stages are shown in Figure \ref{fig:network}. The APA-Net is a cyclic adversarial network based on \citet{zhu2017unpaired} with the addition of anatomy content consistency regularization. As illustrated in Figure \ref{fig:network}, APA-Net adapts images between two domains: the conventional CT domain $\mathbb{A}$ and the CBCT/MR domain $\mathbb{B}$. The anatomy consistency regularization ensures organ and tumor content information are not lost during the unpaired domain adaptation process, thus critical for training a robust segmenter in domain $\mathbb{B}$. Specifically, our APA2Seg-Net contains five networks, including two generators, two discriminators and one segmenter. The generator $G^{A \rightarrow B}$ adapts images from the conventional CT domain to the CBCT/MR domain, the generator $G^{B\rightarrow A}$ adapts the inverse way, the discriminator $D_{B}$ identifies real CBCT or the adapted ones from $G^{A \rightarrow B}$, the discriminator $D_{A}$ identifies real conventional CT or the adapted ones from $G^{B \rightarrow A}$, and the segmenter $M_B$ predicts the segmentation $\hat{S_B}$ on adapted image from generator $G^{A \rightarrow B}$. There are two training paths in our APA2Seg-Net. Path A first adapts conventional CT images $I_A$ to $\hat{I_B}$ in the CBCT/MR domain through $G^{A \rightarrow B}$. Then, $\hat{I_B}$ is adapted back to the conventional CT domain as $\hat{I_A}$ through $G^{B \rightarrow A}$. In parallel, $\hat{I_B}$ is also feed into segmenter $M_B$ to generate segmentation prediction $\hat{S_B}$. Similarly, path B first adapts CBCT/MR images $I_B$ to $\hat{I_A}$ in the conventional CT domain through $G^{B \rightarrow A}$. Then, $\hat{I_A}$ is adapted back to the CBCT/MR domain as $\hat{I_B}$ through $G^{A \rightarrow B}$. 

Training supervision comes from five sources: \\
\noindent\textbf{(a)} adversarial loss $\mathcal{L}_{adv}$ utilizes discriminators to classify if adapted image belong to specific domain. The adversarial objective aims to encourage $G$ to generate adapted images that are indistinguishable to the discriminators. Two adversarial losses are introduced to train generators and discriminators:
\begin{align}
    \mathcal{L}_{adv}(G^{A \rightarrow B},D_{B},I_B,\hat{I_B})&=\E_{I_B\sim\mathbb{B}}\left[ \log{D_{B}(I_B)} \right] \nonumber \\
    &+ \E_{I_A\sim\mathbb{A}}\left[ \log{(1-D_{B}(G^{A \rightarrow B}(I_A)))} \right] \\
    \mathcal{L}_{adv}(G^{B \rightarrow A},D_{A},I_A,\hat{I_A})&=\E_{I_A\sim\mathbb{A}}\left[ \log{D_{A}(I_A)} \right] \nonumber \\
    &+ \E_{I_B\sim\mathbb{B}}\left[ \log{(1-D_{A}(G^{B \rightarrow A}(I_B)))} \right]
\end{align}

\noindent\textbf{(b)} cycle-consistency loss $\mathcal{L}_{cycle}$ constrains the image that returns to the original domain after passing through two generators to have minimal alternation to image content, such that a compound of two generators should be an identity mapping:
\begin{align}
    \mathcal{L}_{cycle}=\E_{I_A\sim\mathbb{A}}\left[ \Vert G^{B \rightarrow A}(G^{A \rightarrow B}(I_A))-I_A \Vert_2^2 \right] \nonumber \\
    + \E_{I_B\sim\mathbb{B}}\left[ \Vert G^{A \rightarrow B}(G^{B \rightarrow A}(I_B))-I_B \Vert_2^2 \right]
\end{align}

\noindent\textbf{(c)} segmentation loss $\mathcal{L}_{seg}$ on the segmentation prediction from image $\hat{I_B}$. The segmentation prediction should be consistent with the ground truth label from the conventional CT domain $\mathbb{A}$: 
\begin{align}
    \mathcal{L}_{seg}=\E_{I_A\sim\mathbb{A}}\left[ 1 - \frac{2 | M_B(G^{A \rightarrow B}(I_A)) \cap S_A |}{|M_B(G^{A \rightarrow B}(I_A))| + |S_A|}  \right]
\end{align}

\noindent\textbf{(d)} identity loss $\mathcal{L}_{idt}$ regularizes the generators to be near an identity mapping when real samples of the target domain are provided. For example, if a given image looks like it is from the target domain, the generator should not map it into a different image. Therefore, the identity loss is formulated as:
\begin{align}
    \mathcal{L}_{idt} = \E_{I_B\sim\mathbb{B}}\left[ \Vert G^{A \rightarrow B}(I_B)-I_B \Vert_2^2 \right] \nonumber \\
    + \E_{I_A\sim\mathbb{A}}\left[ \Vert G^{B \rightarrow A}(I_A)-I_A \Vert_2^2 \right]
\end{align}

\begin{figure*}[htb!]
\centering
\includegraphics[width=0.98\textwidth]{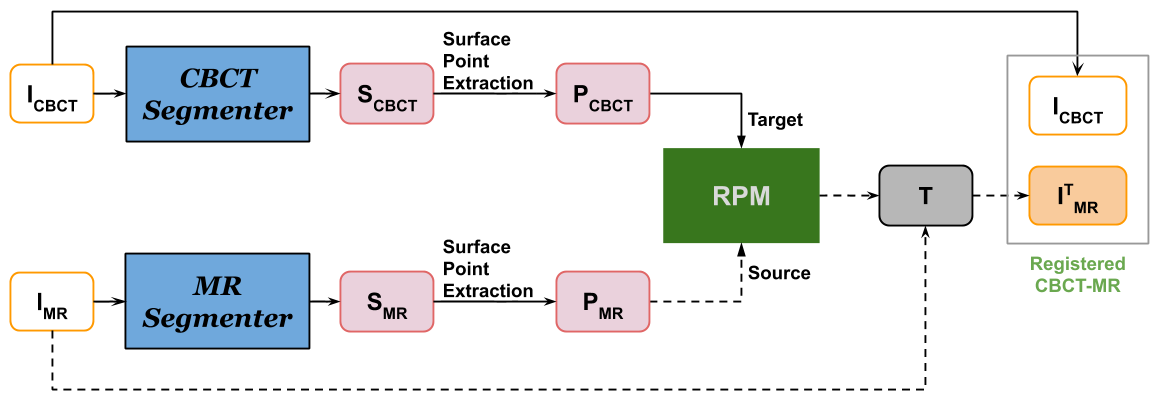}
\caption{Our anatomy-guided multimodal registration pipeline. The segmenters are obtained from APA2Seg-Net in Figure \ref{fig:network}.}
\label{fig:flow_reg}
\end{figure*}

\noindent\textbf{(e)} anatomy-preserving loss $\mathcal{L}_{AP}$ enforces the anatomical content is preserved before and after adaptation. Unlike conventional CycleGAN \citep{zhu2017unpaired} that does not use direct content constraint, we use both the MIND loss \citep{yang2020unsupervised,heinrich2012mind} and correlation coefficient loss \citep{ge2019unpaired} to preserve the anatomy in our unpaired domain adaptation process:
\begin{equation}
    \mathcal{L}_{AP} = \lambda_{cc} \mathcal{L}_{cc} + \lambda_{md} \mathcal{L}_{md}
\end{equation}
The first term $\mathcal{L}_{cc}$ is the correlation coefficient loss, and is formulated as:
\begin{align}
    \mathcal{L}_{cc} = \E_{I_A\sim\mathbb{A}}\left[
    \frac{Cov(G^{A \rightarrow B}(I_A), I_A)}{\sigma_{G^{A \rightarrow B}(I_A)} \; \sigma_{I_A}} \right] \nonumber \\
    + \E_{I_B\sim\mathbb{B}}\left[
    \frac{Cov(G^{B \rightarrow A}(I_B), I_B)}{\sigma_{G^{B \rightarrow A}(I_B)} \; \sigma_{I_B}} \right]
\end{align}
where $Cov$ is the variance operator and $\sigma$ is the standard deviation operator. The second term $\mathcal{L}_{md}$ is the MIND loss \citep{yang2020unsupervised}, and is formulated as:
\begin{align}
    \mathcal{L}_{md} = \E_{I_B\sim\mathbb{B}}\left[ \Vert F(G^{A \rightarrow B}(I_A))-F(I_A) \Vert_1 \right] \nonumber \\
    + \E_{I_A\sim\mathbb{A}}\left[\Vert F(G^{B \rightarrow A}(I_B))-F(I_B) \Vert_1 \right]
\end{align}
where $F$ is a modal-independent feature extractor defined by:
\begin{equation}
    F_x(I) = \frac{1}{Z} exp \left( - \frac{K_x(I)}{V_x(I)} \right)
\end{equation}
where $K_x(I)$ is a distance vector of image patches around voxel $x$ with all the neighborhood patches within a non-local region in image $I$. $V_x(I)$ is the local variance at voxel x in image $I$. Here, dividing $K_x(I)$ with $V_x(I)$ aims to reduce the influence of image modality and intensity range, and $Z$ is a normalization constant to ensure that the maximum element of $F_x$ equals to $1$. In our anatomy-preserving loss, the weight parameters are set to $\lambda_{cc}=1$ and $\lambda_{md}=1$ to achieve balanced training. The anatomy-preserving loss ensures the adaptation only alter the appearance of image while maintaining the anatomical content, such that segmentation network $M_B$ can be trained correctly to recognize the anatomical content in the adapted image. 

Finally, the overall objective is a weighted combination of all loss listed above:
\begin{align}
    \mathcal{L}_{all} = \lambda_{1} \mathcal{L}_{cycle} + \lambda_{2} \mathcal{L}_{adv} + \lambda_{3} \mathcal{L}_{seg} + \lambda_{4} \mathcal{L}_{idt} + \lambda_{5} \mathcal{L}_{AP} 
\end{align}
where weight parameters are set to $\lambda_{1} = 10$ and $\lambda_{2} = \lambda_{3} = \lambda_{4} = \lambda_{5} = 1$ to achieve a balanced training and near-optimal performance according to our hyper-parameter search. The sub-networks' details are shown in Figure \ref{fig:network_details}. Specifically, we use a decoder-encoder network with 9 residual bottleneck for our generators, a 3-layer CNN for our discriminators. Our segmenter is a 5-level UNet with concurrent SE module \cite{roy2018concurrent} concatenated to each level's output. 

\subsection{Anatomy-guided Multimodal Registration}

\begin{figure}[htb!]
\centering
\includegraphics[width=0.49\textwidth]{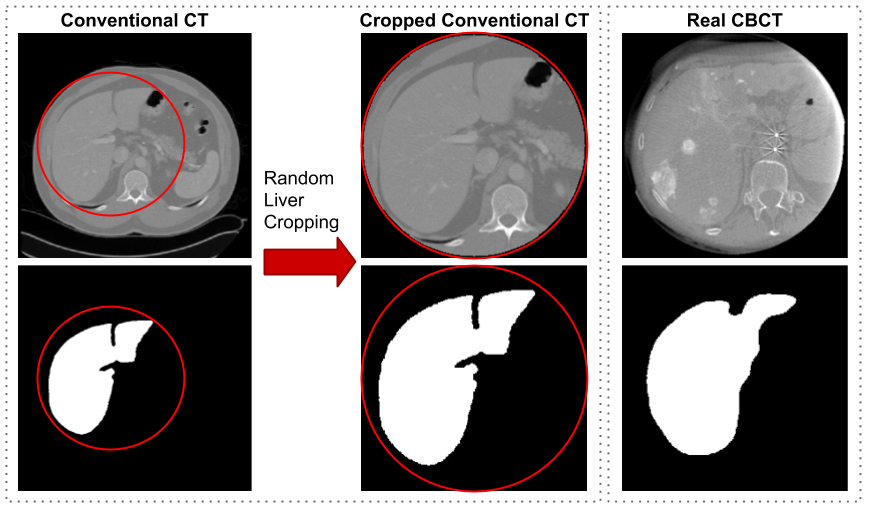}
\caption{Generation of limited FOV CT from conventional CT as input for our APA2Seg-Net.}
\label{fig:crop_ct}
\end{figure}

\begin{figure*}[htb!]
\centering
\includegraphics[width=0.99\textwidth]{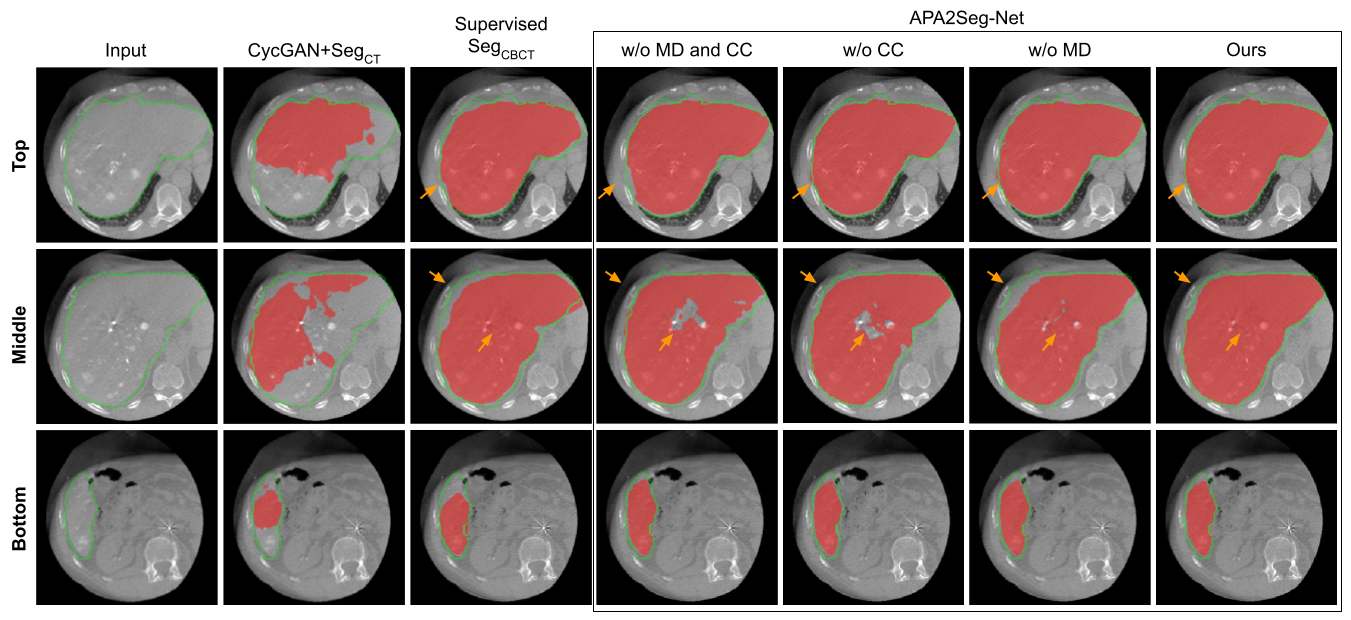}
\caption{Comparison of CBCT segmentation at different liver latitudes. Red mask: liver segmentation prediction. Green contour: liver segmentation ground truth. Results on APA2Seg-Net with or without CC loss and MIND loss are shown in the box.}
\label{fig:seg_cbct}
\end{figure*}

\begin{figure*}[htb!]
\centering
\includegraphics[width=0.99\textwidth]{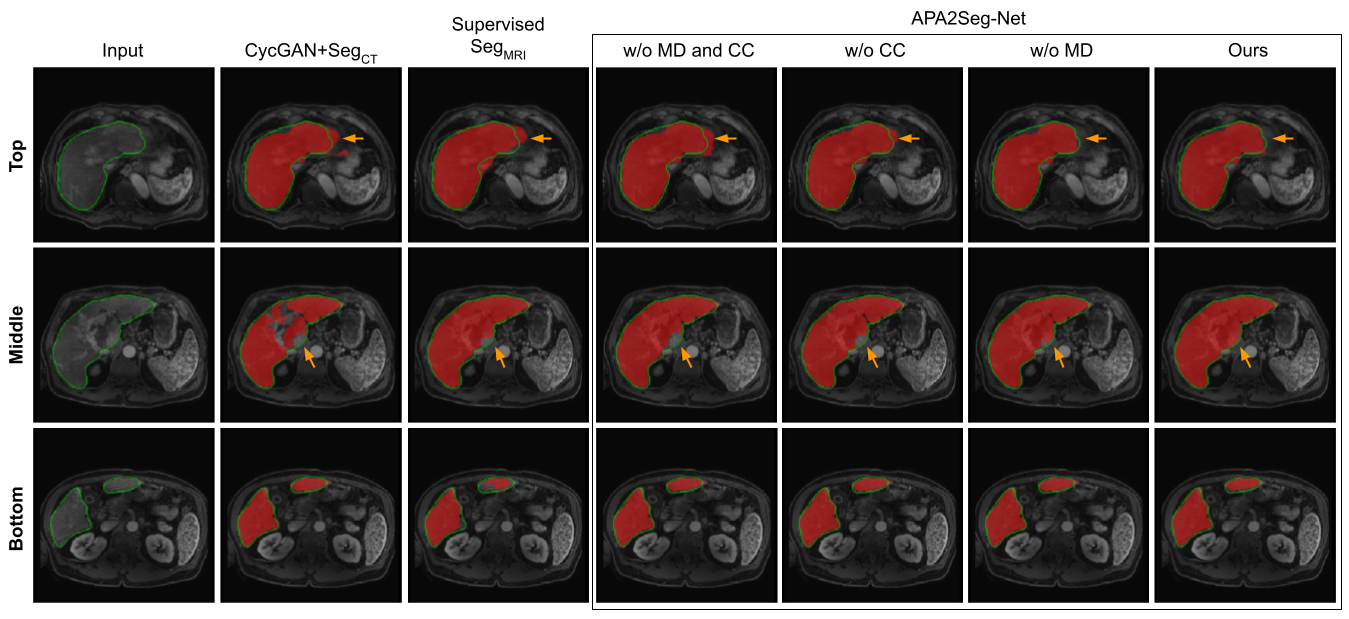}
\caption{Comparison of MRI segmentation at different liver latitudes. Red mask: liver segmentation prediction. Green contour: liver segmentation ground truth. Results on APA2Seg-Net with or without CC loss and MIND loss are shown in the box.}
\label{fig:seg_mri}
\end{figure*}

\begin{table*} [htb!]
\footnotesize
\centering
\caption{Quantitative comparison of CBCT and MRI segmentation results using DSC and ASD(mm). Best results are marked in \textcolor{red}{red}. \underline{Underline} means supervised training with ground truth segmentation on the target domain, i.e. CBCT or MRI segmentation. The negative sign '-' means without the corresponding loss component in our APA2Seg-Net.}
\label{tab:seg}
    \begin{tabular}{|c|c|c||c|c|c|c|c|}
        \hline
        \textbf{CBCT}        & CycGAN+Seg\textsubscript{CT}        & \underline{Seg\textsubscript{CBCT}}  & Ours-Idt           & Ours-MD-CC         & Ours-CC            & Ours-MD            & Ours                                 \\
        \hline
        Median DSC           & $0.685$                             & $0.882$                              & $0.877$            & $0.870$            & $0.878$            & $0.887$            & \textcolor{red}{$0.903$}             \\
        \hline
        Mean$\pm$Std DSC     & $0.695\pm0.092$                     & $0.874\pm0.035$                      & $0.873\pm0.056$    & $0.862\pm0.051$    & $0.871\pm0.051$    & $0.882\pm0.038$    & \textcolor{red}{$0.893\pm0.034$}     \\
        \hline
        Median ASD           & $10.144$                            & $9.190$                              & $6.863$            & $7.289$            & $5.918$            & $6.476$            & \textcolor{red}{$5.882$}             \\
        \hline
        Mean$\pm$Std ASD     & $10.697\pm2.079$                    & $10.742\pm4.998$                     & $6.948\pm2.138$    & $7.459\pm2.769$    & $5.971\pm1.823$    & $6.086\pm1.415$    & \textcolor{red}{$5.886\pm1.517$}     \\
        \hline \hline
        \textbf{MRI}         & CycGAN+Seg\textsubscript{CT}        & \underline{Seg\textsubscript{MRI}}   & Ours-Idt           & Ours-MD-CC         & Ours-CC            & Ours-MD            & Ours                                 \\
        \hline
        Median DSC           & $0.907$                             & $0.907$                              & $0.913$            & $0.915$            & $0.917$            & $0.916$            & \textcolor{red}{$0.918$}             \\
        \hline
        Mean$\pm$Std DSC     & $0.900\pm0.044$                     & $0.859\pm0.102$                      & $0.914\pm0.028$    & $0.912\pm0.029$    & $0.917\pm0.026$    & $0.916\pm0.025$    & \textcolor{red}{$0.921\pm0.022$}     \\
        \hline
        Median ASD           & $1.632$                             & $2.838$                              & $1.619$            & $1.681$            & $1.498$            & $1.532$            & \textcolor{red}{$1.491$}             \\
        \hline
        Mean$\pm$Std ASD     & $2.328\pm2.070$                     & $3.660\pm2.522$                      & $1.932\pm1.303$    & $1.916\pm1.142$    & $1.976\pm1.311$    & $2.047\pm1.382$    & \textcolor{red}{$1.860\pm1.099$}     \\
        \hline
    \end{tabular}
\end{table*}

Using our APA2Seg-Net trained by conventional CT with liver segmentation, we can obtain CBCT and MR segmenters. Then, the segmenters are deployed in our anatomy-guided multimodal registration to guide the Robust Point Matching (RPM) machine to predict the transformation between MR and CBCT images. The registration pipeline is shown in Figure \ref{fig:flow_reg}. The CBCT and MR segmenters from APA2Seg-Net predict the CBCT and MR segmentations. Then, we extract the surface points from the CBCT and MR segmentations and input them into the RPM.

RPM is a point-based registration framework based on deterministic annealing and soft assignment of correspondences between point sets \citep{gold1998new}, which is robust to point outliers. Specifically, given two point sets $\mathcal{M}$ and $\mathcal{S}$, RPM aims to find the affine transformation $T$ that best relates the two point sets. Reformulating the transform $T$ into transformation matrix $A$ and translation vector $t$ form, we have $T(m) =  A m + t$ where $A$ is composed of scale, rotation, horizontal shear, and vertical shear parameters, denoted as $a,\theta,b$, and $c$, respectively. The registration cost function can be written as:
\begin{equation}
    \mathcal{C} = \sum_{j=1}^N \sum_{i=1}^M \mu_{ij} || s_j - (A m_i + t) ||^2 + \alpha (a^2 + b^2 + c^2) - \beta \sum_{j=1}^N \sum_{i=1}^M \mu_{ij}
\end{equation}
where $\mu$ is the point match matrix with $u_{ij}=1$ if point $m_i$ corresponds to point $s_j$ and $u_{ij}=0$ otherwise. The first term minimizes the distance between point sets, and the second term constrains the transformation to avoid large numbers or dramatic transformations. The third term biases the cost toward stronger point correlation by decreasing the cost function. Then, the cost function can be iteratively solved by the soft assignment algorithm. The soft assignment between point sets allows point registration with exclusion of the outlying points and avoids local minima, which fits the problem of CBCT-MR registration well, since the liver is often partially occluded in CBCT due to limited FOV. In addition, for the purpose of intraprocedural registration, RPM provides high-speed registration as it is based on points. The generated transformation $T$ is then applied to original MR image to created registered CBCT-MR pair, which provides better visualization of tumor during image-guided intervention. 

In our implementation, we extract the liver surface points from CBCT and MR segmentation for RPM. Other types of point features from segmentation can also be used in RPM, such as landmarks and skeletons.

\begin{figure*}[htb!]
\centering
\includegraphics[width=1.00\textwidth]{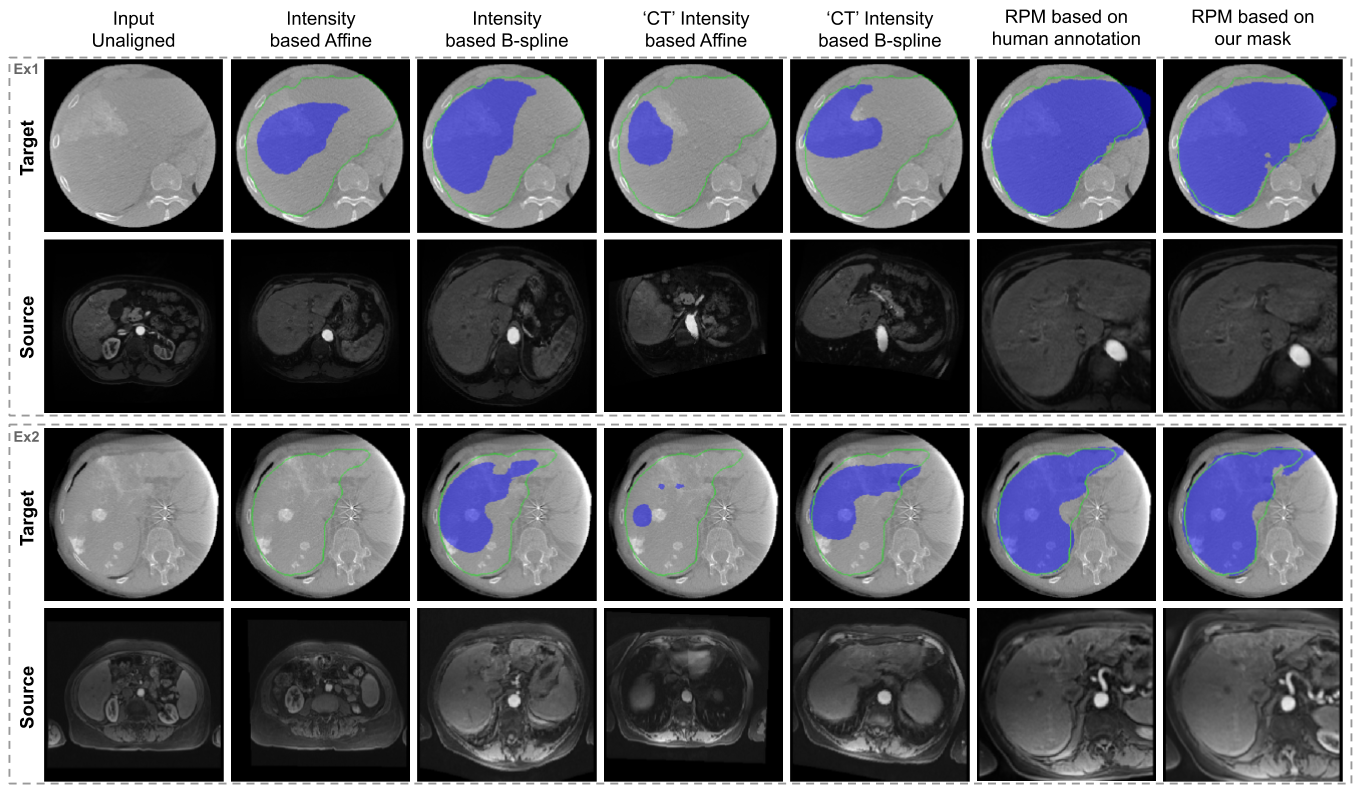}
\caption{CBCT (target) and MRI (source) registration results. Deformation fields are applied on ground truth MRI liver mask and overlaid on CBCT images (blue). Ground truth CBCT liver mask (green contour) is overlaid on CBCT images as well. `CT' Intensity based Affine means intensity-based affine registration based on CT images translated from CBCT and MR using APA-Net. `CT' Intensity based BSpline means intensity-based BSpline registration based on CT images translated from CBCT and MR using APA-Net.}
\label{fig:reg}
\end{figure*}

\section{Experimental Results}
\subsection{Data and Setup}
In the conventional CT domain, we collected 131 and 20 CT volumes with liver segmentation from LiTS \citep{bilic2019liver} and CHAOS \citep{kavur2020chaos}, respectively. In the CBCT/MR domain, we collected 16 in-house TACE patients with both intraprocedural CBCT and pre-operative MR for our segmentation and registration evaluations. All the CBCT data were acquired using a Philips C-arm system with a reconstructed image size of $384 \times 384 \times 297$ and voxel size of $0.65 \times 0.65 \times 0.65 mm^3$. The MR data were acquired using different scanners with different spatial resolutions. Thus, we re-sampled all the CBCT, MR and conventional CT to an isotropic spatial resolution of $1 \times 1 \times 1 mm^3$. 

In the CBCT APA2Seg-Net setup, the conventional CT inputs were first randomly cropped in the axial view using a circular spherical mask on the liver region to simulate the limited FOV in CBCT, as demonstrated in Figure \ref{fig:crop_ct}. The circular mask maintains the same cropping geometry observed in CBCT with a radius of $125mm$. As a result, we obtained $13,241$ 2D conventional CT images with liver segmentation, and $3,792$ 2D CBCT images. In the MR APA2Seg-Net setup, both the conventional CT and MR input were zero-padded or cropped to keep an axial FOV of $410 \times 410 mm^2$. As a result, we obtained $13,241$ 2D conventional CT images with liver segmentation, and $1,128$ 2D MR images. All the 2D images were resized to $256 \times 256$ for APA2Seg-Net inputs. With 16 TACE patients in our dataset, we performed four-fold cross-validation with 12 TACE used as training and 4 patients used as testing in each validation. 


\begin{figure}[htb!]
\centering
\includegraphics[width=0.49\textwidth]{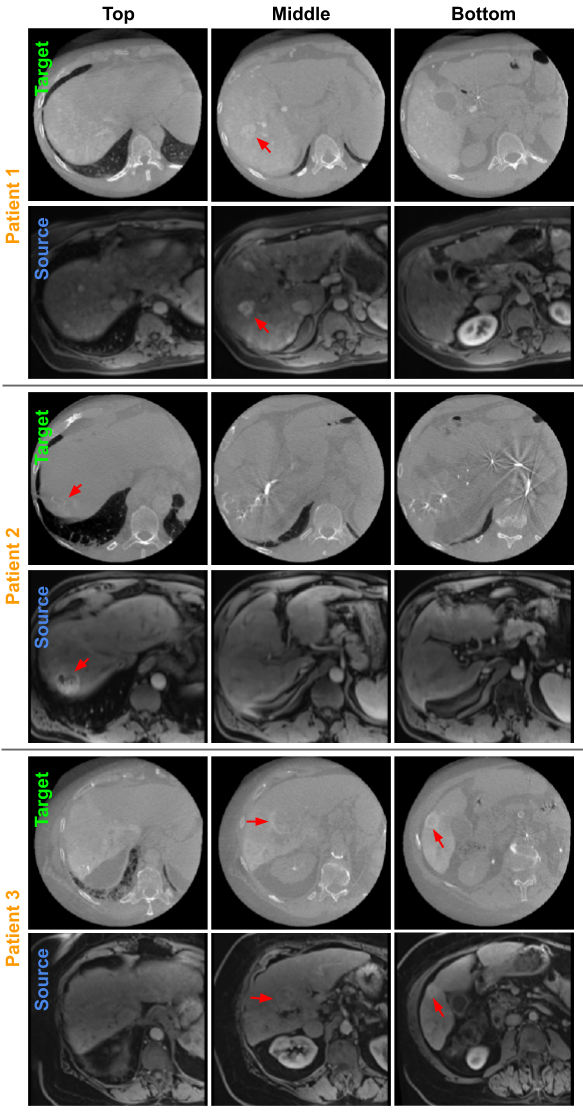}
\caption{Three examples of CBCT (target) and MRI (source) registration results visualized at 3 liver latitudes. RPM registration is performed based on APA2Seg-Net segmentation. Liver tumors are located by red arrows in CBCT and registered MRI.}
\label{fig:reg_additional}
\end{figure}

\subsection{Segmentation Results}
After training, we extracted the segmenters from the APA2Seg-Net for prediction of liver segmentations on both CBCT and MR. For qualitative study, we compared our segmentation performance with: \textbf{i}. CBCT/MR-to-CT CycleGAN concatenated with conventional CT segmenter (CycleGAN+Seg\textsubscript{CT}), where the CT segmenter is trained on conventional CT images with liver annotations (Seg\textsubscript{CT}); \textbf{ii}. APA2Seg-Net without MIND loss and CC loss for anatomy preserving constraint during the training (Ours-MD-CC); \textbf{iii}. APA2Seg-Net with MIND loss only for anatomy preserving constraint during the training (Ours+MD-CC); \textbf{iv}. APA2Seg-Net with CC loss only for anatomy preserving constraint during the training (Ours-MD+CC); \textbf{v}. APA2Seg-Net with both MIND loss and CC loss for anatomy preserving constraint during the training (Ours+MD+CC); and \textbf{vi}. the segmenter trained on target domain images with limited liver annotations (Supervised Seg\textsubscript{CBCT} / Seg\textsubscript{MRI}).

Qualitative comparison of CBCT segmentation results are shown in Figure \ref{fig:seg_cbct}. As we can see, CBCT in TACE suffers from limited FOV, metal artifacts, and low CNR. CycleGAN+Seg\textsubscript{CT} is non-ideal because it requires adapting the input CBCT to conventional CT first, and the segmentation relies on the translated image quality. However, the unpaired and unconstrained adaption from CBCT to CT is difficult as it consists of metal artifact removal and liver boundary enhancement. The multi-stage inference in CycleGAN+Seg\textsubscript{CT} aggregates the prediction error into the final segmentation. On the other hand, our APA2Seg-Net with anatomy-preserving constraint and one-stage inference mechanism achieved significantly better CBCT liver segmentation results. We found combining CC loss and MIND loss for our anatomy-preserving constraint in APA2Seg-Net yields the best results. We also found our identity loss that helps maintain the target domain feature during the adaptation process provides us better segmentation performance. Furthermore, compared to the segmenters trained on target domains using relatively limited annotation data (2844 2D images), our APA2Seg-Net trained from large-scale conventional CT data (13,241 2D images) can provide slightly better segmenters. Qualitative comparison of MR segmentation results are illustrated in Figure \ref{fig:seg_mri}. Similar observations can be found in the MR segmentation results.

Dice Similarity Coefficient (DSC) and Average Symmetric Surface Distance (ASD) were used to evaluate the quantitative segmentation performance. Table \ref{tab:seg} summarizes the quantitative comparison of CBCT and MR segmentation results. As we can see, our APA2Seg-Net achieved the best CBCT and MR segmentation in terms of DSC and ASD, indicating the best overall liver segmentation. 


\subsection{Registration Results}
With the CBCT and MR segmenters extracted from APA2Seg-Nets, we integrate the segmenters into our anatomy-guided multimodal registration pipeline for registering MR to CBCT. The CBCT and MR liver segmentation from segmenters are inputted into RPM to generate the transformation parameters. For qualitative studies, we first compared our registration results with classical previous works of intensity-based affine registration and intensity-based B-spline registration \citep{wyawahare2009image,maes1997multimodality}. We also compared our registration results with intensity-based affine/B-spline registration based on CT images translated from CBCT and MR using APA-Net - similar to the idea in \citet{arar2020unsupervised}. Two examples are illustrated in Figure \ref{fig:reg}. The ground truth (GT) CBCT liver mask (green) and the transformed GT MR liver mask (blue) are overlaid on the CBCT image to qualitatively evaluate the registration performance. As we can observe, neither intensity-based registration methods can correctly estimate the MR transformation, while our anatomy-guided registration, as demonstrated in the last column of Figure \ref{fig:reg}, can more accurately map the MR to CBCT images. Compared to the RPM registration based on ground truth liver segmentations, our anatomy-guided registration based on APA2Seg-Net's segmenter provides similar registration performance. Additional registration results using our method are shown in Figure \ref{fig:reg_additional}.

For quantitative registration evaluation, we first evaluated the averaged error of transformation parameters where the transformation from human annotation based RPM registration is used as ground truth. Based on 3D affine transformation equation:
\begin{align}
T= &
\begin{bmatrix}
   1 & 0 & 0 & \Delta x \\
   0 & 1 & 0 & \Delta y \\
   0 & 0 & 1 & \Delta z \\
   0 & 0 & 0 & 1
\end{bmatrix}
\begin{bmatrix}
   s_x & 0   & 0   & 0 \\
   0   & s_y & 0   & 0 \\
   0   & 0   & s_z & 0 \\
   0   & 0   & 0   & 1
\end{bmatrix}
\begin{bmatrix}
   1      & h_{xy} & h_{xz} & 0 \\
   h_{yx} & 1      & h_{yz} & 0 \\
   h_{zx} & h_{zy} & 1      & 0 \\
   0      & 0      & 0      & 1
\end{bmatrix} \nonumber \\
= & 
\begin{bmatrix}
   s_x          & s_x h_{xy}   & s_x h_{xz}   & \Delta x \\
   s_y h_{yx}   & s_y          & s_y h_{yz}   & \Delta y \\
   s_z h_{zx}   & s_z h_{zy}   & s_z          & \Delta z \\
   0            & 0            & 0            & 1
\end{bmatrix},
\end{align}
twelve 3D affine transformation parameters were evaluated: $s_x$, $s_y$, $s_z$ are the scaling factors on the x,y,z directions, $h_{xy}$, $h_{xz}$, $h_{yx}$, $h_{yz}$, $h_{zx}$, $h_{zy}$ are parameters that control the shear transformation, and $\Delta x$, $\Delta y$, and $\Delta z$ are the translation on the x,y,z directions. The average errors of the parameters are reported in Table \ref{tab:reg_para}. As we can observe, our registration method achieves the least errors in estimating transformation parameters. 


\begin{table} [htb!]
\footnotesize
\centering
\caption{Quantitative comparison of average transformation parameter errors. \underline{Underline} means supervised trained model using ground truth segmentation on the target domain.}
\label{tab:reg_para}
    \begin{tabular}{|c|c|c|c||c|}
        \hline
        Names                & Intst-Affine             & 'CT'-Intst-Affine        & RPM(Ours)                    & RPM(\underline{Seg})       \\
        \hline
        $s_x$                & $3.23$                   & $3.33$                   & $0.23$                       & $0.33$                     \\
        \hline
        $s_y$                & $0.51$                   & $0.62$                   & $0.07$                       & $0.15$                     \\
        \hline
        $s_z$                & $0.62$                   & $0.67$                   & $0.06$                       & $0.26$                     \\
        \hline
        $h_{xy}$             & $0.19$                   & $0.67$                   & $0.04$                       & $0.09$                     \\
        \hline 
        $h_{xz}$             & $0.32$                   & $0.65$                   & $0.04$                       & $0.08$                     \\
        \hline 
        $h_{yx}$             & $0.22$                   & $1.39$                   & $0.12$                       & $0.21$                     \\
        \hline 
        $h_{yz}$             & $0.29$                   & $0.43$                   & $0.07$                       & $0.15$                     \\
        \hline 
        $h_{zx}$             & $0.68$                   & $0.71$                   & $0.25$                       & $0.72$                    \\
        \hline 
        $h_{zy}$             & $0.28$                   & $0.33$                   & $0.09$                       & $0.39$                     \\
        \hline 
        $\Delta x$           & $79.85$ mm               & $73.76$ mm               & $25.13$ mm                   & $65.71$ mm                  \\
        \hline 
        $\Delta y$           & $64.48$ mm               & $38.44$ mm               & $15.05$ mm                   & $27.03$ mm                  \\
        \hline 
        $\Delta z$           & $96.12$ mm               & $161.32$ mm              & $22.74$ mm                   & $52.54$ mm                  \\
        \hline 
    \end{tabular}
\end{table}

\begin{table*} [htb!]
\footnotesize
\centering
\caption{Quantitative comparison of CBCT-MRI registration results using DSC and ASD(mm). Best and second best results are marked in \textcolor{red}{red} and \textcolor{blue}{blue}, respectively. \underline{Underline} means supervised trained model using ground truth segmentation on the target domain. IMG-BSpline and IMG-Affine mean intensity-based BSpline registration and affine registration, respectively. `CT'-IMG-BSpline means intensity-based BSpline registration based on CT images translated from CBCT and MR using APA-Net.}
\label{tab:reg}
    \begin{tabular}{|c|c|c|c|c|c||c|}
        \hline
                             & IMG-BSpline               & IMG-Affine               & `CT'-IMG-BSpline         & RPM(Human Seg)                    & RPM(Ours)                            & RPM(\underline{Seg})       \\
        \hline
        Median DSC           & $0.366$                   & $0.156$                  & $0.332$                  & \textcolor{red}{$0.852$}          & \textcolor{blue}{$0.848$}            & $0.769$                    \\
        \hline
        Mean$\pm$Std DSC     & $0.401\pm0.168$           & $0.168\pm0.049$          & $0.304\pm0.182$          & \textcolor{red}{$0.853\pm0.054$}  & \textcolor{blue}{$0.844\pm0.029$}    & $0.755\pm0.101$            \\
        \hline
        Median ASD           & $25.225$                  & $40.365$                 & $28.424$                 & \textcolor{red}{$4.921$}          & \textcolor{blue}{$5.853$}            & $9.347$                    \\
        \hline
        Mean$\pm$Std ASD     & $25.016\pm8.809$          & $40.682\pm7.187$         & $33.645\pm13.511$        & \textcolor{red}{$5.095\pm1.934$}  & \textcolor{blue}{$5.629\pm0.909$}    & $8.649\pm2.818$            \\
        \hline 
    \end{tabular}
\end{table*}

\begin{figure}[htb!]
\centering
\includegraphics[width=0.49\textwidth]{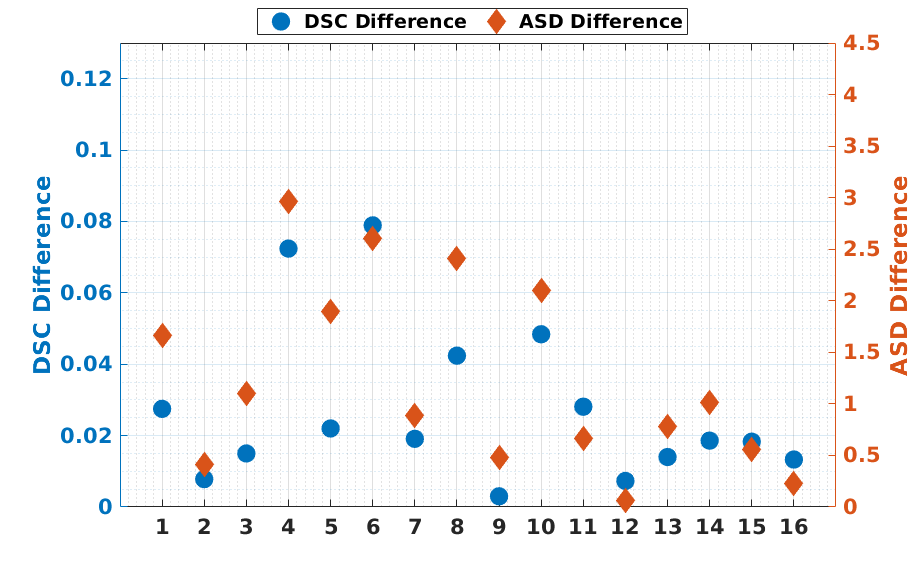}
\caption{Plots of all 16 patients' DSC and ASD differences between RPM registration based on human annotation and RPM registration based on our segmentation.}
\label{fig:diff_reg_casebycase}
\end{figure}

To further validate our registration, we computed the DSC and ASD metrics between the human annotated CBCT liver segmentation and the transformed human annotated MR liver segmentation using transformation generated from different methods. The results are summarized in Table \ref{tab:reg}. Please note that due to the limited FOV of CBCT, the liver mask in CBCT is often truncated while the liver mask in MR is intact. Therefore, the upper limit/gold standard of the metrics are not DSC=1 and ASD=0, but assumed to be the registration results based human annotated liver segmentation. As we can observe from the table, our segmenter-based RPM registration achieved a mean DSC of $0.847$ that is comparable to the human annotation based RPM registration with a mean DSC of $0.853$. Our anatomy-guided method is also significantly better than the intensity-based methods. In Figure \ref{fig:diff_reg_casebycase}, we visualize the case-by-case DSC/ASD differences between our method and the human annotation based registration. Our method can achieve similar registration performance across all 16 cases as compared to the human annotation based registration. A maximal DSC difference less than 0.07 and a maximal ASD difference less than 3mm can be observed. Furthermore, we compared our APA2Seg-Net segementer-based RPM registration to the target domain supervised segmenter-based RPM registration in Table \ref{tab:reg}. We found that our method also outperforms the supervised method that requires annotations on the target domains with the difference significant at $p<0.05$ for both DSC and ASD.

\section{Conclusion and Discussion}
In this work, we proposed an anatomy-guided registration framework by learning segmentation without target modality ground truth. Specifically, we developed an APA2Seg-Net to learn CBCT and MR segmenters without ground truth, which are then plugged into the anatomy-guided registration pipeline for mapping MR to CBCT. We overcame three major difficulties in multimodal image registration. First, we proposed an anatomy-based registration framework that utilizes point clouds of the segmented anatomy, instead of relying on multimodal image intensity which may have significant distribution differences. To obtain robust segmenters of target modality without ground truth, we proposed a segmentation network training scheme without using target modality ground truth, which mitigates the manual annotation requirement on the target modality. Then, we also proposed to use RPM-based point registration that is robust to partially occluded view (point outliers), a scenario commonly observed in TACE and other image-guided intervention procedures. 

We demonstrated the successful application on TACE, in which pre-operative diagnostic MR are registered to intraprocedural CBCT for guiding TACE procedures. Firstly, our method achieved the superior segmentation performance even when compared to the fully supervised methods that requires annotations on the target domains. As annotating new domain data, i.e. intraprocedural CBCT, is not a clinical routine and is time-consuming, one may only obtain limited amount of labeled data for supervised training on the target domain. Thus, it cannot provide sufficient data variability for generating a robust model. On the other hand, our APA2Seg-Net utilizing large-scale conventional CT dataset offers much larger data variability, thus achieved superior segmentation performance even without using ground truth annotations from the target domain. Then, given the more robust segmenters from APA2Seg-Net, our registration pipeline based on these segmenters and RPM can also offer superior registration performance. In Table \ref{tab:reg}, our method is able to reduce the ASD between MR and CBCT liver segmentation from ~4 cm based on previous intensity-based affine registration to ~0.5cm, and reduce the translation difference from 9.6cm based on previous intensity-based affine registration to ~2cm, as demonstrated in Table \ref{tab:reg_para}. With our method, the registration errors now fall within a more acceptable range. This allows MR to be more accurately registered to CBCT, reinforcing the utility of MR-derived information within the clinical TACE image guidance environment.

The presented work also has potential limitations. First of all, the CBCT segmentation performance is far from perfect with a mean DSC of 0.893. In our current APA2Seg-Net implementation, only 2D networks were considered in this study since the amount of training data is not large enough to train a robust 3D network. However, the proposed APA2Seg-Net can be extended to 3D with the expense of higher GPU memory consumption and longer computation time, which would potentially provide better segmentation results if a large amount of 3D training scans is available. As a matter of fact, \citet{zhang2018translating,cai2019towards} had demonstrated the promising results from 3D synthesis and segmentation. On the other hand, our APA2Seg-Net is an open framework with flexibility in network components. While we used 2D Res-Net/Patch GAN/concurrent-SE-UNet as our generator/discriminator/segmenter, we do not claim optimality of the combination for segmentation. Other image segmentation networks, such as attention UNet \citep{oktay2018attention}, multi-scale guided attention network \citep{sinha2020multi}, and adversarial image-to-image network \citep{yang2017automatic}, could be deployed in our framework and might yield better segmentation performance on different applications. Secondly, as the segmentation is imperfect from our APA2Seg-Net, it leads to difference between registration based on our segmentation and human segmentation. However, the impact of imperfect CBCT/MR segmentation is mitigated through our RPM based registration. As we can observe from Table \ref{tab:reg} and Figure \ref{fig:reg}, the human segmentation based registration is very close to the APA2Seg-Net segmentation-based registration in terms of qualitative visualization and quantitative comparison ($<0.01$ in terms of Dice). Thirdly, we considered affine transformation in our CBCT-MR liver registration, and incorporating non-rigid registration could potentially provide more accurate internal structure mappings. However, liver in CBCT is often truncated due to limited FOV. Therefore, using non-rigid registration would lead to incorrect matching on the truncated boundary. Future works on incorporating non-rigid registration while rejecting the point outliers outside of FOV is needed. Lastly, we evaluated the registration performance on the entire liver, while registration performance on other important landmarks, such as tumor location, is not included here. We will evaluate other landmark's alignment in our future clinical feasibility studies.

The design of our anatomy-guided registration framework by learning segmentation without ground truth also suggests several interesting topics for future studies. First of all, our method could be adapted to other multimodal registration tasks that conventional registration techniques are not applicable, such as MR-Ultrasound (US) registration for neurosurgery \citep{rivaz2014automatic} and prostate interventions \citep{hu2012mr}, where occluded FOV and intensity inhomogeneity is often observed in US. There are several public datasets containing MR brain tumor and MR prostate with ground truth segmentation \cite{simpson2019large}, which make it possible to adapt our method to these applications. More specifically, we could use APA2Seg-Net to obtain the US and MR segmenters, which are then embedded into our anatomy-guided registration pipeline for real-time MR-US alignments. Similar to our idea in registration, \citet{sultana2019deformable} recently proposed a prostate US-PET/CT registration algorithm based on segmentation for dose planning \citet{drean2016interindividual}, in which our APA2Seg-Net could potentially provide the US and PET/CT prostate segmenter as well. Secondly, our method could also be adapted to landmark-based registration tasks. While anatomy-guided registration framework based on segmentation is demonstrated in this work, the segmenter in APA2Seg-Net could be replaced with a detector for learning keypoint detection without ground truth on target domain. Then, the keypoint detector could also be embedded into our anatomy-guided registration pipeline for keypoint based alignments. 

In summary, we proposed an anatomy-guided registration framework by learning segmentation without target modality ground truth based on APA2Seg-Net. We demonstrated the successful application on intraprocedural CBCT-MR liver registration. In the future, we will assess the tumor-of-interest's registration accuracy and evaluate the clinical impact of real-time intraprocedural MR-CBCT liver registration.

\section*{Acknowledgments}
This work was supported by funding from the National Institutes of Health/ National Cancer Institute (NIH/NCI) under grant number R01CA206180. BZ was supported by the Biomedical Engineering Ph.D. fellowship from Yale University.

\section*{Credit authorship contribution statement }
\textbf{Bo Zhou}: Conceptualization, Methodology, Software, Visualization, Validation, Formal analysis, Writing original draft. 
\textbf{Zachary Augenfeld}: Methodology, Writing - review and editing.
\textbf{Julius Chapiro}: Writing - review and editing.
\textbf{S. Kevin Zhou}: Writing - review and editing.
\textbf{Chi Liu}: Writing - review and editing, Supervision.
\textbf{James S. Duncan}: Conceptualization, Methodology, Writing - review and editing, Supervision.

\bibliographystyle{model2-names.bst}\biboptions{authoryear}
\bibliography{refs}

\end{document}